\documentclass[journal]{IEEEtran}
\usepackage[utf8]{inputenc}
\usepackage[T1]{fontenc}
\usepackage{graphicx}
\usepackage{subfigure}
\usepackage{amsmath}
\usepackage{xcolor}
\usepackage{siunitx}
  
\begin{document}

\title{Dynamic EEE Coalescing: Techniques and Bounds}

\author{Sergio~Herrería-Alonso,
  Miguel~Rodríguez-Pérez,~\IEEEmembership{Senior~Member,~IEEE,}
  Manuel~Fernández-Veiga,~\IEEEmembership{Senior~Member,~IEEE,}
  and~Cándido~López-García%
  \thanks{The authors are with the Department of Telematics
    Engineering, University of Vigo, 36310 Vigo, Spain (e-mail:
    sha@det.uvigo.es).}}

\maketitle

\begin{abstract}
  Frame coalescing is one of the most efficient techniques to manage
  the low power idle (LPI) mode supported by Energy Efficient Ethernet
  (EEE) interfaces. This technique enables EEE~interfaces to remain in
  the LPI~mode for a certain amount of time upon the arrival of the
  first frame (time-based coalescing) or until a predefined amount of
  traffic accumulates in the transmission buffer (size-based
  coalescing). This paper provides new insights on the practical
  efficiency limits of both coalescing techniques. In particular, we
  derive the fundamental limits on the maximum energy savings
  considering a target average frame delay. Additionally, we present
  new open-loop adaptive variants of both time-based and size-based
  coalescing techniques. These proposals dynamically adjust the length
  of the sleeping periods in accordance with actual traffic conditions
  to reduce energy consumption while keeping the average delay near a
  predefined value simultaneously. Analytical and simulation results
  show that the energy consumption of both proposals is comparable to
  the fundamental limits. Consequently, we recommend the usage of the
  time-based algorithm in most scenarios because of its simplicity as
  well as its ability to bound the maximum frame delay at the same
  time.
\end{abstract}


\section{Introduction}
\label{sec:intro}

\IEEEPARstart{C}{urrent} Ethernet interfaces have the ability to save
power by entering a low power idle (LPI) mode whenever there is no
traffic to transmit.  This LPI mode, defined in the Energy Efficient
Ethernet~(EEE) amendment to the Ethernet standard~\cite{802.3az}, only
needs around~\SI{10}{\percent} of the energy used during normal
operation. However, transitions to and from the LPI mode consume
energy and take some time to complete, so proper care has to be placed
to decide when and for how long to use this LPI mode. Possibly to spur
innovation, the standard does not offer guidelines as for how to
employ this mode, leaving the task to devise efficient LPI controllers
to hardware designers.  Probably, the most popular family of governing
algorithms is the one based on \emph{frame coalescing} (also known as
\emph{burst transmission})~\cite{christensen10:_the_road_to_eee,
  reviriego10:_burst_tx_eee, herreria11:_oppor_ether}. These
algorithms strive to minimize energy usage by staying in the LPI mode
until a significant amount of traffic is ready for
transmission. Unfortunately, this has the undesired side-effect of
increasing frame delay, so a careful balance between traffic delay and
energy consumption is required.

Coalescing algorithms can be fundamentally subdivided into two
complementary categories based on the signal used to abandon the
LPI~mode: time-based coalescers and size-based ones. The first kind
determines the amount of coalesced traffic in the LPI mode indirectly,
firing a timer when the first frame is queued for transmission. When
the timer expires, the interface returns to the normal operating
mode. Size-based coalescers, on the contrary, exit the LPI mode when a
predefined amount of traffic accumulates. In both cases, the proper
tuning of the timer value, or the queue threshold, is critical to get
good performance without suffering excessive QoS
degradation~\cite{choi12:_tradeoff,herreria12:_optimal_conf_eee}. A
further complication stems from the fact that actual traffic
characteristics influence the coalescing parameter tuning, so there is
no single value that performs well enough for any traffic
load~\cite{herreria12:_bounded_energy}.

This paper provides new insights on the practical efficiency limits of
coalescing techniques. In particular, we derive the fundamental limits
on the maximum energy savings considering a target average frame
delay. Our second contribution is a couple of new open-loop dynamic
coalescing algorithms: a time-based one and another from the
size-based family. Both algorithms dynamically adjust their
corresponding coalescing parameters in accordance with actual traffic
conditions to reduce energy consumption while keeping the average
delay near a predefined value. To asses their relative goodness, we
also compare both algorithms against the practical bounds under
different traffic conditions. The obtained results show that if the
target delay is higher than a few microseconds, the energy consumption
of both proposals closely approximates the fundamental
limits. Finally, as a result of this comparison, we provide guidelines
for the selection of the most convenient algorithm in accordance with
the allowable delay characteristics. In any case, we can anticipate
that in most scenarios, the time-based algorithm is to be preferred
because of its simplicity as well as its ability to bound the maximum
frame delay at the same time.

The rest of the paper is organized as follows. Related work is
reviewed in Sect.~\ref{sec:related}. Section~\ref{sec:model} presents
the energy consumption and delay models on which we will build our
proposals. To facilitate the understanding of the dynamic techniques
and the computation of the practical bounds, we summarize in
Sect.~\ref{sec:t_coalescing} and~\ref{sec:s_coalescing} the Poisson
models developed in~\cite{herreria12:_gig1_model} for both time-based
and size-based coalescing algorithms. In Sect.~\ref{sec:d_coalescing}
the new adaptive versions of both coalescing algorithms are
presented. Then, in Sect.~\ref{sec:bound}, we find a lower bound for
the energy consumed under the constraint of a target average
delay. The proposed dynamic techniques are mathematically analyzed and
evaluated through simulation experiments in Sect.~\ref{sec:eval} and,
based on the obtained results, we provide some guidelines for their
application in Sect.~\ref{sec:recommend}. Finally, in
Sect.~\ref{sec:conclusions} we summarize the main conclusions of this
work.

\section{Related Work}
\label{sec:related}

\subsection{EEE Coalescing}

Ordinarily, EEE~interfaces enter the LPI~mode every time the
transmission buffer gets empty and, if no coalescing is applied, they
resume normal operation as soon as new traffic arrives. Unfortunately,
this simple algorithm does not usually provide satisfying results
since it triggers an excessive number of mode transitions and a great
amount of energy is wasted on them~\cite{reviriego11:evaleee}. To
reduce the frequency of transitions, coalescing algorithms enable
EEE~interfaces to remain in the LPI~mode until a significant amount of
traffic is ready for
transmission~\cite{christensen10:_the_road_to_eee,
  reviriego10:_burst_tx_eee, herreria11:_oppor_ether}. Certainly,
coalescing frames into bursts extends idle periods but, sadly, also
increases traffic delay. If the coalescing algorithm is configured
with a long timer duration (or a high queue threshold), frames may
suffer excessively large delays. On the contrary, if the coalescing
parameter is configured with a too low value, only modest energy
savings will be achieved. There is, therefore, a trade-off between
energy consumption and frame
delay~\cite{choi12:_tradeoff,herreria12:_optimal_conf_eee}. Moreover,
traffic characteristics affect the coalescing parameter tuning, so
there is no single value that performs well enough for any traffic
load. Consequently, coalescers should dynamically tune the coalescing
parameter according to actual traffic conditions to achieve the
desired performance~\cite{herreria12:_bounded_energy}.

The research community has already provided some dynamic tuning
algorithms over the last years. In~\cite{herreria12:_bounded_energy}
the authors tune a size-based coalescer to obtain a predefined energy
efficiency target. The tuning of a time-based coalescer (and a
size-based coalescer) to meet a target average delay is discussed
in~\cite{chatzipapas16:_static_dyn_coalescing}
(in~\cite{chatzipapas13:_static_dyn_coalescing,
  chatzipapas16:_mg1_gigabit_eee}, resp.). All the aforementioned
approaches rely on a feedback loop for tuning the coalescing parameter
so, when traffic conditions are themselves dynamic, there is a
convergence period. As is usual for closed-loop systems, the speed of
convergence is controlled with a feedback gain parameter that must be
carefully configured to guarantee system stability as well as a rapid
response to varying traffic conditions. In contrast, we propose new
open-loop dynamic algorithms that do not rely on any feedback signal
to adapt to actual traffic conditions and maintain system stability.

Finally, \cite{pan17:_power_delay_tradeoff}~provides rules to select
the appropriate queue threshold and timer duration values to comply
with the average delay requirement when using both size-based and
time-based coalescing
jointly. \cite{guo16:base_station_sleeping_control} also obtains
similar results for the case of sleeping base stations using
size-based coalescing. Unfortunately, these papers do not provide a
dynamic method for adjusting the coalescing parameters, they give no
indication of how and when to estimate the traffic variables required
to compute the optimal coalescing values and the computation of the
optimal queue threshold is excessively complicated to be carried out
in real-time.

\subsection{EEE Analytical Models}

Many analytical models evaluating EEE~behavior have been developed in
recent years. Some of them do not consider coalescing and just assume
that EEE~interfaces awake by the first frame
arrival~\cite{marsan11:_analytic_eee, larrabeiti11:_10gb_opt_eth,
  bolla14:_802.3az_model}. Other works just provide models for the
energy consumed using coalescing with Poisson
traffic~\cite{herreria11:_power_savin_model_burst_trans,
  herreria11:_how_effic_is_energ_effic_ether}. More interesting are
those models addressing the energy-delay trade-off. For instance,
\cite{mostowfi12:_coalescer}~presents a deterministic model for
size-based coalescers while \cite{akar13:_anal_timer_based}~models
time-based coalescers for Poisson arrivals. Most models analyze the
general case that considers the joint use of time-based and size-based
coalescing algorithms, both for
1000BASE-T~interfaces~\cite{chatzipapas13:_static_dyn_coalescing,
  chatzipapas16:_mg1_gigabit_eee} and for 10GBASE-T
ones~\cite{herreria12:_optimal_conf_eee, herreria12:_gig1_model,
  pan17:_power_delay_tradeoff, kim13:_math_802.3az,
  meng17:_understanding_burst}.

In this paper, we focus on \SI{10}{\giga b\per\second} (and faster)
EEE~interfaces,\footnote{In \SI{10}{\giga b\per\second} (and faster)
  EEE~interfaces, frame arrivals cannot interrupt transitions from
  active to the LPI~mode. In addition, mode transitions can take place
  in each link direction in an independent way.} so we build on the
GI/G/1 model proposed in~\cite{herreria12:_gig1_model} for 10GBASE-T
interfaces since it provides precise but easy-to-use expressions for
the computation of the energy consumption and the average frame delay
when using coalescing.

\section{Energy Consumption and Delay Models}
\label{sec:model}

This paper builds on the analytical model developed
in~\cite{herreria12:_gig1_model}, so we assume that frame arrivals
follow a general distribution with independent interarrival times
$I_n$, $n=1,2,\ldots$, and average arrival rate~$\lambda$ while the
service times $S_n$, $n=1,2,\ldots$, are a set of random variables
with equal distribution function and mean service
rate~$\mu$. Furthermore, we assume an utilization
$\rho=\lambda/\mu<1$, thus assuring system stability.

\begin{figure*}[t]
  \centering
  \resizebox{0.9\textwidth}{!}{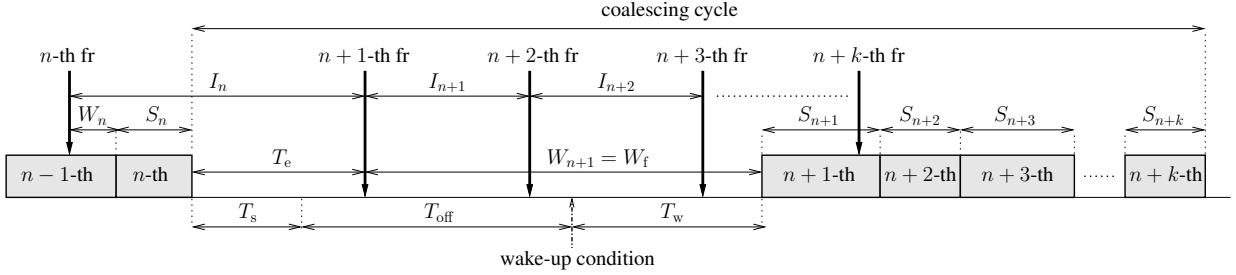}
  \caption{Example of EEE coalescing operations.}
  \label{fig:busy-cycle}
\end{figure*}

Figure~\ref{fig:busy-cycle} shows an example of EEE~operations when
using coalescing. In particular, the example shows a complete
coalescing cycle during which $k$~frames are received and sent (from
the $(n+1)$-th to the $(n+k)$-th frame). To maximize energy savings,
the EEE~interface is put to sleep as soon as its transmission buffer
gets empty and, after a short transition of length~$T_{\mathrm{s}}$,
it enters the LPI~mode. The interface remains in the LPI~mode for a
period of random length~$T_{\mathrm{off}}$. Once the wake-up condition
dictated by the coalescing technique is met, the interface abandons
the LPI~mode and, after a transition of length~$T_{\mathrm{w}}$, it
starts transmitting the frames received while sleeping. Note that,
after the interface is put to sleep, there exists a period of random
length~$T_{\mathrm{e}}$ with no frames in the transmission buffer
(with no job to do). Also note that all the received frames in the
coalescing cycle have to wait some amount of time in the transmission
queue before they can be transmitted. We denote by $W_n$ the queuing
delay experienced by the $n$-th~frame.

\subsection{Energy Consumption}

According to~\cite{herreria12:_gig1_model}, the energy consumed by an
EEE interface compared with that consumed by a power-unaware interface
that does not support the LPI~mode is given by
\begin{equation}
  \label{eq:pow}
  \varphi = 1 - (1 - \varphi_{\mathrm{off}}) (1 - \rho)
  \frac{\overline{T}_{\mathrm{off}}}{\overline{T}_{\mathrm{off}} + T_{\mathrm{s}} +
    T_{\mathrm{w}}}\,,
\end{equation}
where $\varphi_{\mathrm{off}}$ is the ratio of the energy consumed by
the EEE~interface in the LPI~mode to that consumed in the active state
and $\overline{T}_{\mathrm{off}}$ is the average time the
EEE~interface remains in the sleeping mode in each coalescing
cycle. This expression was obtained making the usual assumption that
EEE~interfaces approximately consume the same amount of energy during
transitions as in the active state. Note that, since
$\varphi_{\mathrm{off}}$, $T_{\mathrm{s}}$ and $T_{\mathrm{w}}$ are
intrinsic characteristics of EEE~interfaces, we only need to compute
the average length of sleeping periods, $\overline{T}_{\mathrm{off}}$,
to obtain their energy consumption. We assume that
$\varphi_{\mathrm{off}}=0.1$, $T_{\mathrm{s}}=$ \SI{2.88}{\us} and
$T_{\mathrm{w}}=$ \SI{4.48}{\us} as the standard states for typical
10GBASE-T interfaces.

\subsection{Average Queuing Delay}

It is shown in~\cite{herreria12:_gig1_model} that the average queuing
delay when using coalescing is given by the following expression:
\begin{equation}
  \label{eq:W}
  \overline{W} = \overline{W}_{\mathrm{0}} + \frac{\overline{W_{\mathrm{f}}^2}
    -\overline{T_{\mathrm{e}}^2}}{2(\overline{W}_{\mathrm{f}}+\overline{T}_{\mathrm{e}})}
  - \frac{\lambda \mathrm{cov}(W,I)}{1-\rho}\,, 
\end{equation}
where $W_{\mathrm{f}}$ is the random queuing delay experienced by the
first frame in the coalescing cycle, $T_{\mathrm{e}}$ is the empty
period, that is, the random time elapsed since the interface is put to
sleep until the first frame arrives, $\mathrm{cov}(W,I)$ is the
covariance between the queuing time experienced by a frame and the
time until the next frame arrives and
\begin{equation}
  \label{eq:W0}
  \overline{W}_{\mathrm{0}} = \frac{\lambda^2(\sigma_I^2 +
    \sigma_S^2) + (1-\rho)^2}{2\lambda(1-\rho)}
\end{equation}
is a fixed term independent of the coalescing algorithm, with
$\sigma_I^2$ and $\sigma_S^2$ being, respectively, the variances of
interarrival and service times.

In the following sections this general model will be particularized
for both time-based and size-based coalescing techniques with Poisson
traffic. Although it is well-known that frame arrivals are not
exponentially distributed~\cite{paxson95:_wide_area_traffic}, they are
often modeled as Poisson processes for analytic simplicity. In
addition, Poisson traffic can serve as a reasonable approximation to
real traffic in sub-second time
scales~\cite{karagiannis04:_nonst_poiss_view_of_inter_traff} and to
aggregated traffic in Internet core
links~\cite{vishwanath09:_how_poisson_is_tcp_traffic}.

\section{Time-based Coalescing}
\label{sec:t_coalescing}

We summarize in this section the model developed in
\cite{herreria12:_gig1_model} for the time-based coalescing technique
with Poisson traffic.

\subsection{Average Length of Sleeping Periods}

With time-based coalescing, the sleeping period ends when a predefined
timer of duration~$V$ started upon the arrival of the first frame
expires. Therefore, the interface clearly remains
$T_{\mathrm{e}}+V-T_{\mathrm{s}}$ seconds in the LPI~mode. Assuming
$V>T_{\mathrm{s}}$ to guarantee that every transition to sleep makes
the interface spend some time in the LPI~mode, the average length of
sleeping periods is given by
\begin{equation}
  \label{eq:T_off_time_based}
  \overline{T}_{\mathrm{off}}^{\mathrm{tb}} =
  \int_{0}^{\infty} \! (t+V-T_{\mathrm{s}}) f_{T_{\mathrm{e}}}(t) \, \mathrm{d}t\,,
\end{equation}
where $f_{T_{\mathrm{e}}}(t)$ is the probability density function of
the duration of empty periods. Due to the memoryless property of
Poisson traffic, empty periods and interarrival times are identically
distributed, so $f_{T_{\mathrm{e}}}(t) = f_{I_n}(t) = \lambda
\mathrm{e}^{-\lambda t}, t \ge 0$. Then, solving
integral~\eqref{eq:T_off_time_based}, we get
\begin{equation}
  \label{eq:T_off_time_based_poisson}
  \overline{T}_{\mathrm{off}}^{\mathrm{tb}} = 1/\lambda+V-T_{\mathrm{s}}\,.
\end{equation}

\subsection{Average Queuing Delay}

In this case, the first frame in the coalescing cycle will always wait
$V+T_{\mathrm{w}}$ since the sleeping timer is just fired when it
arrives, so $\overline{W}_{\mathrm{f}}=V+T_{\mathrm{w}}$ and
$\overline{W_{\mathrm{f}}^2}=(V+T_{\mathrm{w}})^2$. In addition,
$\mathrm{cov}(W,I) = 0$ since the waiting time of a frame is
independent of the arrival time of the next frame and
$\overline{T}_{\mathrm{e}}=1/\lambda$ (and
$\overline{T_{\mathrm{e}}^2} = \sigma_I^2 +
\overline{T}_{\mathrm{e}}^2 = 1/\lambda^2 + 1/\lambda^2 =
2/\lambda^2$) due to the Poisson memoryless property. Substituting all
these values in~\eqref{eq:W}, we finally get
\begin{equation}
  \label{eq:W_time_based_poisson}
  \overline{W}^{\mathrm{tb}} = \overline{W}_{\mathrm{0}} +
  \frac{\lambda^2(V+T_{\mathrm{w}})^2-2}{2\lambda(1+\lambda(V+T_{\mathrm{w}}))}\,.
\end{equation}

\section{Size-based Coalescing}
\label{sec:s_coalescing}

In this section we summarize the model developed in
\cite{herreria12:_gig1_model} for size-based coalescers with Poisson
traffic.

\subsection{Average Length of Sleeping Periods}

With size-based coalescing, a sleeping interface wakes up when the
number of frames queued in the transmission buffer reaches a
predefined threshold~$Q_{\mathrm{w}}$, so
\begin{equation}
  \label{eq:T_off_size_based}
  \overline{T}_{\mathrm{off}}^{\mathrm{sb}} =
  \int_{T_{\mathrm{s}}}^{\infty} \! (t-T_{\mathrm{s}}) f_{Q_{\mathrm{w}}}(t) \, \mathrm{d}t\,,
\end{equation}
where $f_{Q_{\mathrm{w}}}(t)$ is the probability density function of
the time elapsed since the interface is put to sleep until the arrival
of the $Q_{\mathrm{w}}-$th~frame. With Poisson traffic, all
interarrival times are identically and exponentially distributed, so
the arrival time of the $Q_{\mathrm{w}}-$th~frame is
Erlang-$Q_{\mathrm{w}}$ distributed and
$f_{Q_{\mathrm{w}}}(t)=\lambda^{Q_{\mathrm{w}}} t^{Q_{\mathrm{w}}-1}
\mathrm{e}^{-\lambda t} / (Q_{\mathrm{w}}-1)!$. Then, solving
integral~\eqref{eq:T_off_size_based}, we get
\begin{equation}
  \label{eq:T_off_size_based_poisson}
  \overline{T}_{\mathrm{off}}^{\mathrm{sb}} = \frac{\Gamma
    (Q_{\mathrm{w}}+1,\lambda T_{\mathrm{s}}) - \lambda T_{\mathrm{s}}
    \Gamma (Q_{\mathrm{w}},\lambda T_{\mathrm{s}})}{\lambda \Gamma
    (Q_{\mathrm{w}})}\,,
\end{equation}
where $\Gamma (q,x) = \int_x^{\infty} t^{q-1} \mathrm{e}^{-t}\,
\mathrm{d}t$ is the upper incomplete gamma function and $\Gamma (q) =
\Gamma (q,0)$.

\subsection{Average Queuing Delay}

With this technique, the first frame arriving in the coalescing cycle
will wait for the arrival of the next $Q_{\mathrm{w}}-1\,$frames, so
\begin{equation}
  \label{eq:W_f_size_based}
  \overline{W}_{\mathrm{f}}^{\mathrm{sb}} =
  \int_0^{\infty} \! (t+T_{\mathrm{w}}) f_{Q_{\mathrm{w}}-1}(t) \, \mathrm{d}t\,.
\end{equation}
As just explained, when considering Poisson traffic,
$f_{Q_{\mathrm{w}}-1}(t)$ is the Erlang-$Q_{\mathrm{w}}-1$ probability
density function, so solving this integral, we get
$\overline{W}_{\mathrm{f}}=(Q_{\mathrm{w}}-1)/\lambda+T_{\mathrm{w}}$,
and then $\overline{W_{\mathrm{f}}^2} =
\sigma_{W_\mathrm{f}}^2+(\overline{W}_{\mathrm{f}})^2 =
(Q_{\mathrm{w}}~-~1)\sigma_I^2+((Q_{\mathrm{w}}-1)/\lambda+T_{\mathrm{w}})^2$. In
addition, $\mathrm{cov}(W,I)$ has a positive value since the queuing
delay of the first $Q_{\mathrm{w}}-1$ frames depends on the next
interarrival times. \cite{heyman68:_bounds_single_server_queues}
proves that, in this scenario, $\mathrm{cov}(W,I)$ is given by
\begin{equation}
  \label{eq:cov_W_I}
  \mathrm{cov}(W,I) = \frac{(1 - \rho) (Q_{\mathrm{w}} - 1)
    \sigma_I^2}{Q_{\mathrm{w}} - 1 + \lambda \overline{T}_{\mathrm{e}} }\,.
\end{equation}
Finally, substituting these values in~\eqref{eq:W}, we have
\begin{equation}
  \label{eq:W_size_based_poisson}
  \overline{W}^{\mathrm{sb}} =  \overline{W}_{\mathrm{0}} 
  - \frac{Q_{\mathrm{w}} - 1}{\lambda Q_{\mathrm{w}}}
  + \frac{(Q_{\mathrm{w}}+\lambda T_{\mathrm{w}}-1)^2 + Q_{\mathrm{w}} - 3}{2\lambda(Q_{\mathrm{w}} + \lambda T_{\mathrm{w}})}\,.
\end{equation}

\section{Dynamic Coalescing}
\label{sec:d_coalescing}

As previously shown, both energy consumption and average queuing delay
depend on the value configured for the coalescing parameter ($V$ or
$Q_{\mathrm{w}}$). Consequently, the average queuing delay can be kept
around a desired value~$\tau$ if the coalescing parameter is
dynamically and suitably configured according to actual traffic
conditions. For example, with time-based coalescing,
equating~\eqref{eq:W_time_based_poisson} to~$\tau$ and solving for
$V$, we get that
\begin{equation}
  \label{eq:V_opt}
  V^* = \tau - \overline{W}_{\mathrm{0}} - T_{\mathrm{w}} +
  \frac{1}{\lambda}\sqrt{1+\left(1+\lambda(\tau - \overline{W}_{\mathrm{0}})\right)^2}
\end{equation}
is the timer duration required to reach the average delay~$\tau$.

Similarly, equating~\eqref{eq:W_size_based_poisson} to~$\tau$, we find
that, with size-based coalescing, the coalescing threshold
$Q_{\mathrm{w}}^*$ needed to keep the average delay~$\tau$ must meet
the condition
\begin{IEEEeqnarray}{rCl}
  \label{eq:Q_opt}
  Q_{\mathrm{w}}^{*3}\: && + \left(2\lambda T_{\mathrm{w}}-2\lambda
  (\tau-\overline{W}_{\mathrm{0}})-3\right) Q_{\mathrm{w}}^{*2} \IEEEnonumber\\
  &&+ \left(\lambda^2 T_{\mathrm{w}}^2-2\lambda^2 T_{\mathrm{w}}(\tau-\overline{W}_{\mathrm{0}})-4\lambda T_{\mathrm{w}} \right) Q_{\mathrm{w}}^* \IEEEnonumber\\
  &&+\: 2\lambda T_{\mathrm{w}} = 0,
\end{IEEEeqnarray}
that can be resolved using any of the multiple algebraic methods known
to find the roots of cubic equations.

Therefore, to guarantee a given average delay, adaptive coalescing
techniques should dynamically adjust the coalescing parameter
following the guidelines just provided in this section. We recommend
to make this adjustment each time the transmission buffer gets empty,
just before putting the interface to sleep at the beginning of a new
coalescing cycle. This assures a quick reaction to variable traffic
conditions.

However, note that the computation of the optimal coalescing
parameters requires accurate estimations of the average arrival
rate~$\lambda$ and the $\overline{W}_{\mathrm{0}}$~delay, as shown
in~\eqref{eq:V_opt} and~\eqref{eq:Q_opt}. But the computation of
$\overline{W}_{\mathrm{0}}$ requires, in turn, to estimate the average
service rate (required to compute the utilization~$\rho$) and the
variances of both interarrival and service times, as shown
in~\eqref{eq:W0}. This could certainly hinder the implementation of
the dynamic techniques. Hence, to simplify
$\overline{W}_{\mathrm{0}}$~computation, we propose to assume Poisson
arrivals ($\sigma_I^2=1/\lambda^2$) with deterministic frame sizes
($\sigma_S^2=0$), so that $\overline{W}_{\mathrm{0}}$ can be
approximated as
\begin{equation}
  \label{eq:W0approx}
  \overline{W}_{\mathrm{0}} \approx \frac{1 + (1-\rho)^2}{2\lambda(1-\rho)},
\end{equation}
and it is only necessary to measure the average arrival rate and the
average service rate. These variables can be estimated at the
beginning of each new coalescing cycle, just before adjusting the
corresponding coalescing parameter. For example, the average arrival
rate could be computed just dividing the amount of frames received in
the last coalescing cycle by its duration. Similarly, the average
service rate could be estimated just dividing the nominal rate of the
interface by the average size of the frames received in the last
coalescing cycle.

On the other hand, recall that cubic equation~\eqref{eq:Q_opt} must be
resolved to obtain the optimal queue threshold required to reach the
target delay. This computation can be greatly simplified if we assume
that $Q_{\mathrm{w}}$ is configured with a large value. Under this
condition, the average queuing delay
in~\eqref{eq:W_size_based_poisson} can be approximated as
\begin{equation}
  \label{eq:W_size_based_poisson_approx}
  \overline{W}^{\mathrm{sb}} \approx \overline{W}_{\mathrm{0}} + 
  \frac{Q_{\mathrm{w}}+\lambda T_{\mathrm{w}}-3}{2\lambda},
\end{equation}
and then the optimal queue threshold can be estimated as
\begin{equation}
  \label{eq:Q_opt_approx}
  Q_{\mathrm{w}}^* \approx 2\lambda\left(\tau-\overline{W}_{\mathrm{0}}-\frac{T_{\mathrm{w}}}{2}\right) + 3.
\end{equation}

\begin{figure}[t]
  \centering
  \subfigure[Timer durations.]{
    \includegraphics[width=0.85\columnwidth]{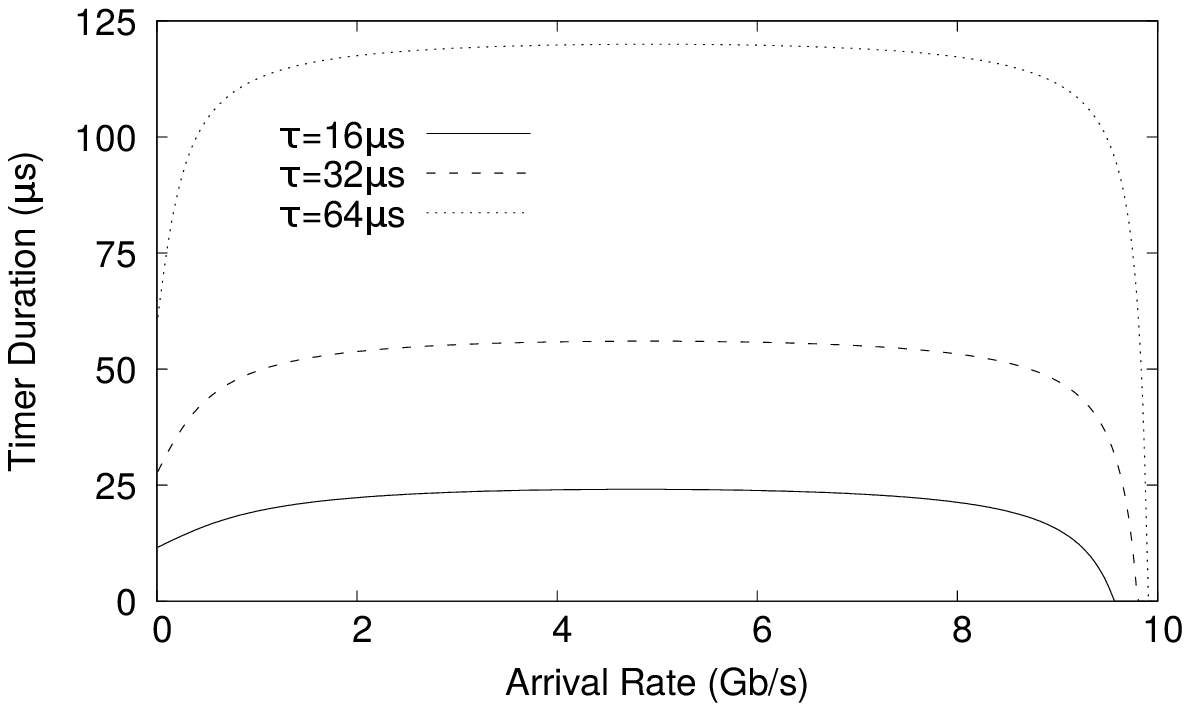}
    \label{fig:timer-duration}
  }
  \subfigure[Queue thresholds.]{
    \includegraphics[width=0.85\columnwidth]{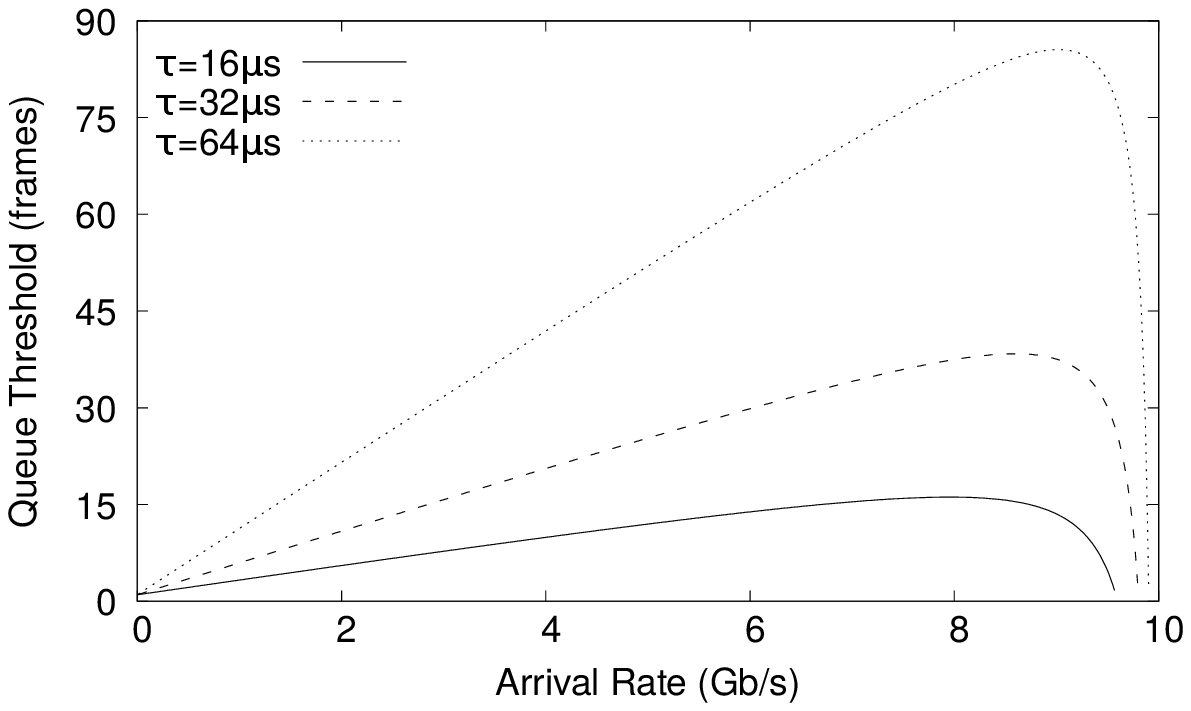}
    \label{fig:queue-threshold}
  }
  \caption{Coalescing parameters required to reach different target
    delays.}
  \label{fig:coal-parameters}
\end{figure}

Figure~\ref{fig:coal-parameters} shows the timer durations and the
queue thresholds computed using~\eqref{eq:V_opt}
and~\eqref{eq:Q_opt_approx} to maintain different average queuing
delays ($\tau=16,\,32$ and \SI{64}{\us}) with $1500$-byte frames. The
queue thresholds obtained solving cubic equation~\eqref{eq:Q_opt} are
not shown in Fig.~\ref{fig:queue-threshold} since they are
indistinguishable from those obtained with
approximation~\eqref{eq:Q_opt_approx}. Note that the optimal
coalescing parameters take invalid values ($V^*<0$,
$Q_{\mathrm{w}}^*<1$) for the highest (and most unlikely) rates. At
very high rates (those higher than $9.5\,$Gb/s), the transmission
buffer fills up due to traffic arriving faster than it is processed
and the average queuing delay, even in the absence of any sleeping
algorithm, will exceed the target delay. So, under these extreme
traffic conditions, the delay constraint cannot be fulfilled and the
interface should remain active without ever entering the LPI~mode thus
preventing the queuing delay from increasing even more. Note that this
is not a shortcoming of the proposed techniques, but simply an
undesired effect of the unavoidable increase on frame delay due to an
excessive load.

\section{Lower Bound for Energy Consumption Given a Target Average Delay}
\label{sec:bound}

Clearly, the more time the EEE~interfaces remain in the sleeping mode,
the less amount of energy they will consume. Therefore, to compute a
lower bound for the energy consumed by EEE~interfaces under the
constraint of a given average delay, we must previously obtain an
upper bound for the average length of sleeping periods
($\overline{T}_{\mathrm{off}}$) under this condition. We will build on
the general model presented in Sect.~\ref{sec:model}. Assuming that
the average queuing delay equals the target value~$\tau$, substituting
$\overline{W_{\mathrm{f}}^2} = \sigma_{W_\mathrm{f}}^2 +
(\overline{W}_{\mathrm{f}})^2$ and $\overline{T_{\mathrm{e}}^2} =
\sigma_{T_\mathrm{e}}^2 + (\overline{T}_{\mathrm{e}})^2$
in~\eqref{eq:W}, and solving for $\overline{W}_{\mathrm{f}}$, we get
that the average queuing delay of the first frame in each coalescing
cycle must hold
\begin{IEEEeqnarray}{rCl}
  \label{eq:W_first_tau}
  && \overline{W}_{\mathrm{f}} = \tau - \overline{W}_{\mathrm{0}} +
  \frac{\lambda \mathrm{cov}(W,I)}{1-\rho} \IEEEnonumber \\
  &&+\: \sqrt{\left(\tau-\overline{W}_{\mathrm{0}}+\frac{\lambda \mathrm{cov}(W,I)}{1-\rho} +
    \overline{T}_{\mathrm{e}}\right)^2 + \sigma_{T_\mathrm{e}}^2 - \sigma_{W_\mathrm{f}}^2}.
\end{IEEEeqnarray}
On the other hand, it can be easily seen from
Fig.~\ref{fig:busy-cycle} that $\overline{W}_{\mathrm{f}} =
T_{\mathrm{s}} + T_{\mathrm{w}} + \overline{T}_{\mathrm{off}} -
\overline{T}_{\mathrm{e}}$, so equating this and
\eqref{eq:W_first_tau}, and solving for $\overline{T}_{\mathrm{off}}$,
we have
\begin{IEEEeqnarray}{rCl}
  \label{eq:T_off_tau}
  && \overline{T}_{\mathrm{off}} = \tau - T_{\mathrm{s}} - T_{\mathrm{w}} - \overline{W}_{\mathrm{0}} +
  \frac{\lambda \mathrm{cov}(W,I)}{1-\rho} + \overline{T}_{\mathrm{e}} \IEEEnonumber \\
  &&+\: \sqrt{\left(\tau-\overline{W}_{\mathrm{0}}+\frac{\lambda \mathrm{cov}(W,I)}{1-\rho} +
    \overline{T}_{\mathrm{e}}\right)^2 + \sigma_{T_\mathrm{e}}^2 - \sigma_{W_\mathrm{f}}^2}.
\end{IEEEeqnarray}
Next we will compute an upper bound for $\overline{T}_{\mathrm{off}}$.
Firstly, note from~\eqref{eq:T_off_tau} that greater sleeping periods
are obtained with longer (and more varying) empty periods. Let the
$n-$th frame be the last frame in a busy cycle. As shown in
Fig.~\ref{fig:busy-cycle}, $T_{\mathrm{e}}=I_n-W_n-S_n$, so,
$\overline{T_{\mathrm{e}}}< 1/\lambda-1/\mu$ because the last frame in
a busy cycle always has to wait some amount of time in the
transmission queue ($W_n>0$). Moreover, $\sigma_{T_\mathrm{e}}^2 =
\sigma_I^2 + \sigma_S^2 + \sigma_{W_n}^2 + 2\mathrm{cov}(W_n,S_n) -
2\mathrm{cov}(W_n,I_n) - 2\mathrm{cov}(I_n,S_n)$, but
$\mathrm{cov}(W_n,S_n)=\mathrm{cov}(I_n,S_n)=0$ since the waiting time
experienced by a frame is independent of its length and it is assumed
that frame lengths do not depend on the arrival process. Thus,
$\sigma_{T_\mathrm{e}}^2 < \sigma_I^2 + \sigma_S^2 + \sigma_{W_n}^2$,
and, since $\sigma_{W_n}^2 < \overline{W_n^2} < \overline{(I_n-S_n)^2}
= \sigma_{I_n-S_n}^2+(\overline{I_n-S_n})^2 = \sigma_I^2 + \sigma_S^2
+ (1/\lambda-1/\mu)^2$, we get that $\sigma_{T_\mathrm{e}}^2 <
2(\sigma_I^2 + \sigma_S^2) + (1/\lambda-1/\mu)^2$.

Additionally, note that longer sleeping periods can be obtained with
greater $\mathrm{cov}(W,I)$~terms, so, to avoid that this covariance
is zero, we assume that the waiting time of a frame depends on the
arrival time of the next frame as is the case with size-based
coalescing. The $\mathrm{cov}(W,I)$ term can be computed by
conditioning on the arrival times of the frames within the coalescing
cycle. Note that, for those frames arriving while the interface is
sleeping, this covariance is $\sigma_I^2$ while for those ones
arriving once the interface has been reactivated, this term is
zero. Averaging over the coalescing cycle, we get that
$\mathrm{cov}(W,I)=\sigma_I^2
\overline{N}_{\mathrm{off}}/\overline{N}$ where
$\overline{N}_{\mathrm{off}}$ is the average number of frames received
while the interface is sleeping and $\overline{N}$ is the average
number of frames served in the whole coalescing cycle. Clearly,
$\overline{N}_{\mathrm{off}}=\lambda (\overline{W}_{\mathrm{f}}-
T_{\mathrm{w}})$. In addition, it is well-known that, in a GI/G/1
system with vacations,
$\overline{N}=\lambda(\overline{W}_{\mathrm{f}}+\overline{T}_{\mathrm{e}})/(1-\rho)$
where $\overline{W}_{\mathrm{f}}+\overline{T}_{\mathrm{e}}$ is the
average vacation time. Therefore, we get that
\begin{equation}
  \label{eq:cov_bound}
  \mathrm{cov}(W,I) = (1-\rho) \frac{\overline{W}_{\mathrm{f}} -
    T_{\mathrm{w}}}{\overline{W}_{\mathrm{f}} +
    \overline{T}_{\mathrm{e}}} \sigma_I^2 \, < \, (1-\rho)\sigma_I^2.
\end{equation}

Then, assuming that $\sigma_{W_\mathrm{f}}^2=0$ (as in the best case
with time-based coalescing) and substituting the upper limits of
$\overline{T}_{\mathrm{e}}$, $\sigma_{T_\mathrm{e}}^2$ and
$\mathrm{cov}(W,I)$ in~\eqref{eq:T_off_tau}, we obtain the following
upper bound for $\overline{T}_{\mathrm{off}}$:
\begin{IEEEeqnarray}{rCl}
  \label{eq:T_off_tau_max}
  && \overline{T}_{\mathrm{off}} < \tau - T_{\mathrm{s}} - T_{\mathrm{w}} - \overline{W}_{\mathrm{0}}
  + \lambda \sigma_I^2 + \frac{1-\rho}{\lambda} + \IEEEnonumber \\
  && \sqrt{\left(\tau - \overline{W}_{\mathrm{0}} + \lambda \sigma_I^2 + \frac{1-\rho}{\lambda}\right)^2
    + 2(\sigma_I^2 + \sigma_S^2) + \left(\frac{1-\rho}{\lambda}\right)^2}\,.
\end{IEEEeqnarray}
  
Finally, this value can be substituted in~\eqref{eq:pow} to obtain a
lower bound for the energy consumed by EEE~interfaces under the
constraint of a given average queuing delay.

\section{Evaluation}
\label{sec:eval}

\begin{figure}[t]
  \centering
  \subfigure[Average length of sleeping periods.]{
    \includegraphics[width=0.85\columnwidth]{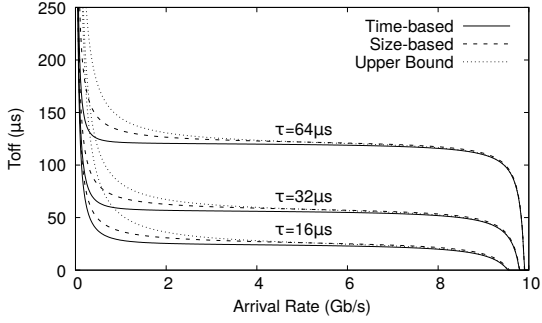}
    \label{fig:poisson-results-toff}
  }
  \subfigure[Energy consumption.]{
    \includegraphics[width=0.85\columnwidth]{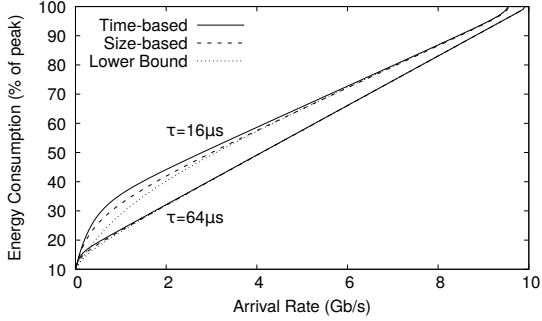}
    \label{fig:poisson-results-energy}
  }
  \caption{Analytical results with Poisson traffic.}
  \label{fig:poisson-results}
\end{figure}

We firstly used the main results obtained in the analytical model to
compute the energy consumed by EEE~interfaces when using dynamic
coalescing with Poisson traffic and \num{1500}-byte frames.
Figure~\ref{fig:poisson-results-toff} shows the average lengths of
sleeping periods for both time-based and size-based coalescing
techniques when the coalescing parameters are configured following the
guidelines provided in Sect.~\ref{sec:d_coalescing}. The upper bound
given by~\eqref{eq:T_off_tau_max} is also shown in the graph. The
results evidence that, at low rates, size-based coalescing achieves
longer sleeping periods, although there is still some room for
improvement. However, from moderate to high rates, the duration of the
sleeping periods obtained with both techniques are practically
identical and match the upper bound. Also note that the average length
of the sleeping periods for the highest rates tends to zero since,
under these extreme conditions, the sleeping algorithm should be
temporarily suspended to prevent the queuing delay from increasing
even more.

Figure~\ref{fig:poisson-results-energy} shows the energy consumed with
the proposed techniques and the lower bound obtained when
substituting~\eqref{eq:T_off_tau_max} in~\eqref{eq:pow}. As expected,
dynamic size-based coalescing achieves greater energy savings since it
provides longer sleeping periods, especially at low rates. However,
note that, for a moderate target delay of just \SI{64}{\us}, the
energy consumed with both techniques is very similar and, remarkably,
very close to the lower bound, even at the lowest rates. And, although
not shown in the graph for clarity, the same occurs for greater target
delays. Therefore, our proposals are able to minimize energy
consumption in practice except when the target delay is configured
with an excessively small value.

\subsection{Comparison with Static Coalescing}

We then compared the proposed schemes with static coalescing
techniques in which the coalescing parameter is configured with a
fixed value regardless of actual traffic conditions as those analyzed
in~\cite{herreria12:_gig1_model, kim13:_math_802.3az,
  meng17:_understanding_burst}. To ensure a fair comparison, the
coalescing parameters of the static coalescers were configured with
the values required to get an average delay equal to the corresponding
target delay when $\lambda=5\,$Gb/s. Thus, after
using~\eqref{eq:V_opt} and~\eqref{eq:Q_opt_approx} to compute the
coalescing parameters required for a target delay of \SI{16}{\us}
(\SI{64}{\us}), we set the timer duration for the static time-based
coalescer to \SI{24}{\us} (\SI{120}{\us}) and the queue threshold for
the static size-based one to $12\,$frames ($52\,$frames).

\begin{figure}[t]
  \centering
  \subfigure[Energy consumption.]{
    \includegraphics[width=0.85\columnwidth]{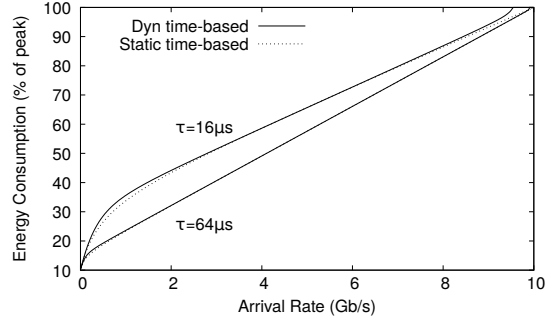}
    \label{fig:comparison-time-based-energy-results}
  }
  \subfigure[Average queuing delay.]{
    \includegraphics[width=0.85\columnwidth]{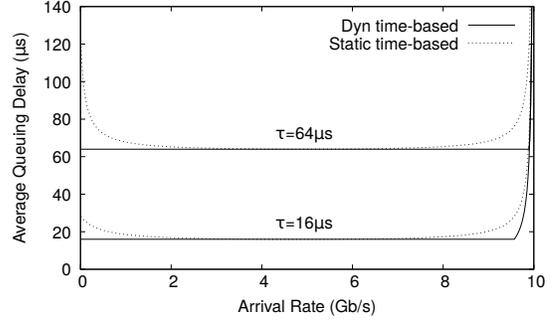}
    \label{fig:comparison-time-based-delay-results}
  }
  \caption{Comparison between dynamic and static time-based coalescing.}
  \label{fig:comparison-time-based-results}
\end{figure}

Figure~\ref{fig:comparison-time-based-results} shows the energy
consumption and the average queuing delay when using both dynamic and
static time-based coalescers. With a \SI{16}{\us} target delay, static
time-based coalescing consumes a little less energy than the dynamic
one at the lowest and the highest loads, but the average delay is
moderately increased as well. If the target delay is increased to
\SI{64}{\us}, then frames still suffer longer delays with static
coalescing at extreme loads but without obtaining significant energy
savings this time.

\begin{figure}[t]
  \centering
  \subfigure[Energy consumption.]{
    \includegraphics[width=0.85\columnwidth]{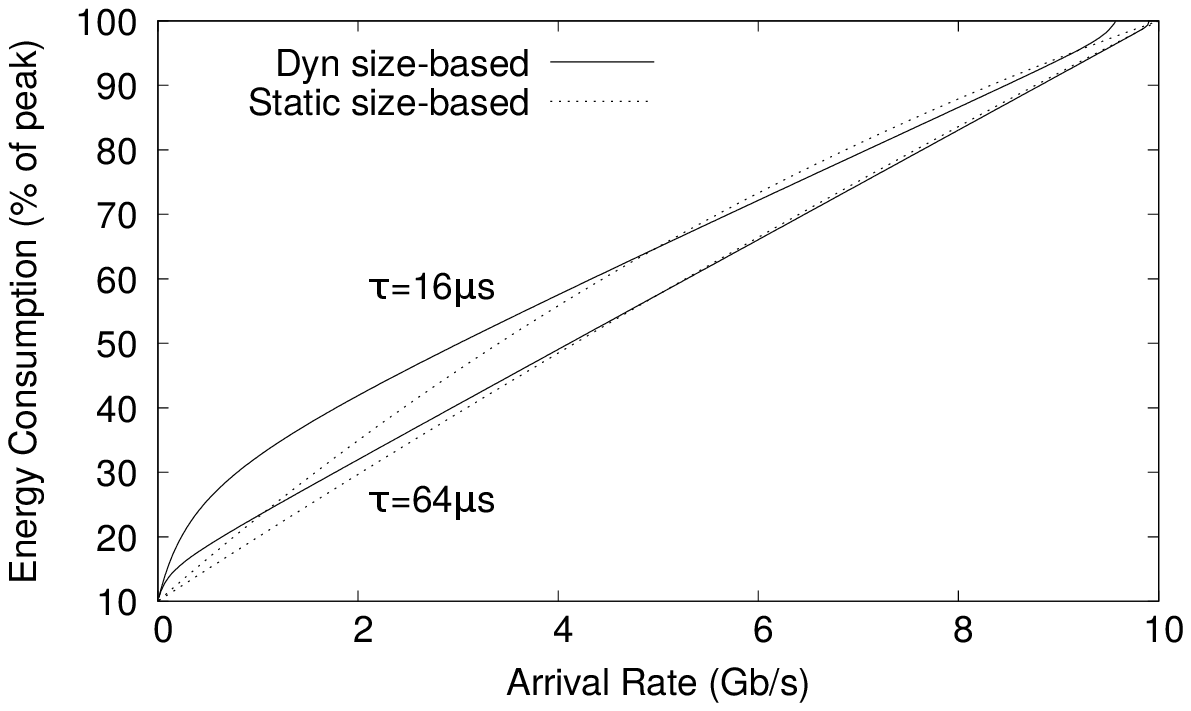}
    \label{fig:comparison-size-based-energy-results}
  }
  \subfigure[Average queuing delay.]{
    \includegraphics[width=0.85\columnwidth]{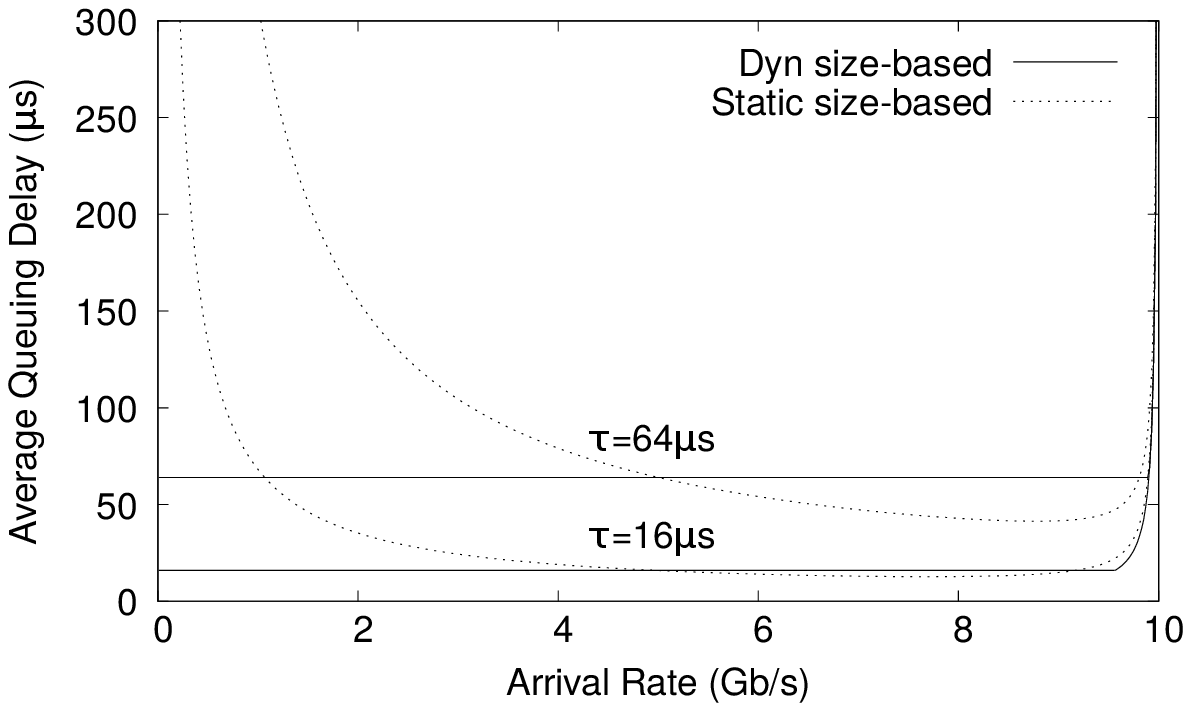}
    \label{fig:comparison-size-based-delay-results}
  }
  \caption{Comparison between dynamic and static size-based coalescing.}
  \label{fig:comparison-size-based-results}
\end{figure}

Figure~\ref{fig:comparison-size-based-results} shows the results
obtained with both dynamic and static size-based coalescers. When
using static size-based coalescing, deviations from the target delay
are huge, especially at the lowest rates where the obtained delays are
just unacceptable since reaching the queue threshold takes an
excessively long time. At high rates, the queue threshold is reached
too soon, so the average delay is below the target value but at the
expense of a small increment in energy consumption.

For acceptable performance, static coalescers need to jointly use both
time-based and size-based techniques. Thus, at rates lower than the
configured rate threshold ($5\,$Gb/s in these experiments), the
sleeping timer will expire before reaching the queue threshold in most
occasions. Therefore, the coalescer will behave as a time-based one
and the queuing delay will be bounded. Conversely, at rates higher
than $5\,$Gb/s, the queue threshold will be likely reached before the
sleeping timer expires, so the coalescer will behave as a size-based
one thus reducing the frame delay at the cost of a slight increase in
energy consumption. In any case, with static coalescing, it is
impossible to maintain the average delay near the target value for all
the possible traffic loads. At low rates, the average delay will
exceed the desired value while, at high rates, the delay will get
excessively reduced, thus wasting some energy needlessly.

\subsection{Dynamic Coalescing with Pareto Traffic and Bimodal Frame Sizes}

We also evaluated the proposed techniques with an open-source in-house
simulator~\cite{herreria15:dualmode_eee_simulator} under more
stringent conditions. In the following simulation experiments, we
considered Pareto interarrival times with shape parameter $\alpha=2.5$
to validate our formulas with self-similar traffic.\footnote{Note that
  Pareto distributions must be characterized with a shape parameter
  $\alpha$ greater than~$2$ to have finite variance. On the other
  hand, the greater the $\alpha$~parameter, the shorter the
  fluctuations, so a value of~$2.5$ is a good trade off to have finite
  variance along with significant fluctuations.} Additionally, to
approximate real Internet traffic, frame sizes were set to follow a
bimodal distribution with 54\% of frames having a size of $100\,$bytes
and the rest with a size of
$1500\,$bytes~\cite{murray12:_network_traffic}.

\begin{figure}[t]
  \centering
  \subfigure[Energy consumption.]{
    \includegraphics[width=0.85\columnwidth]{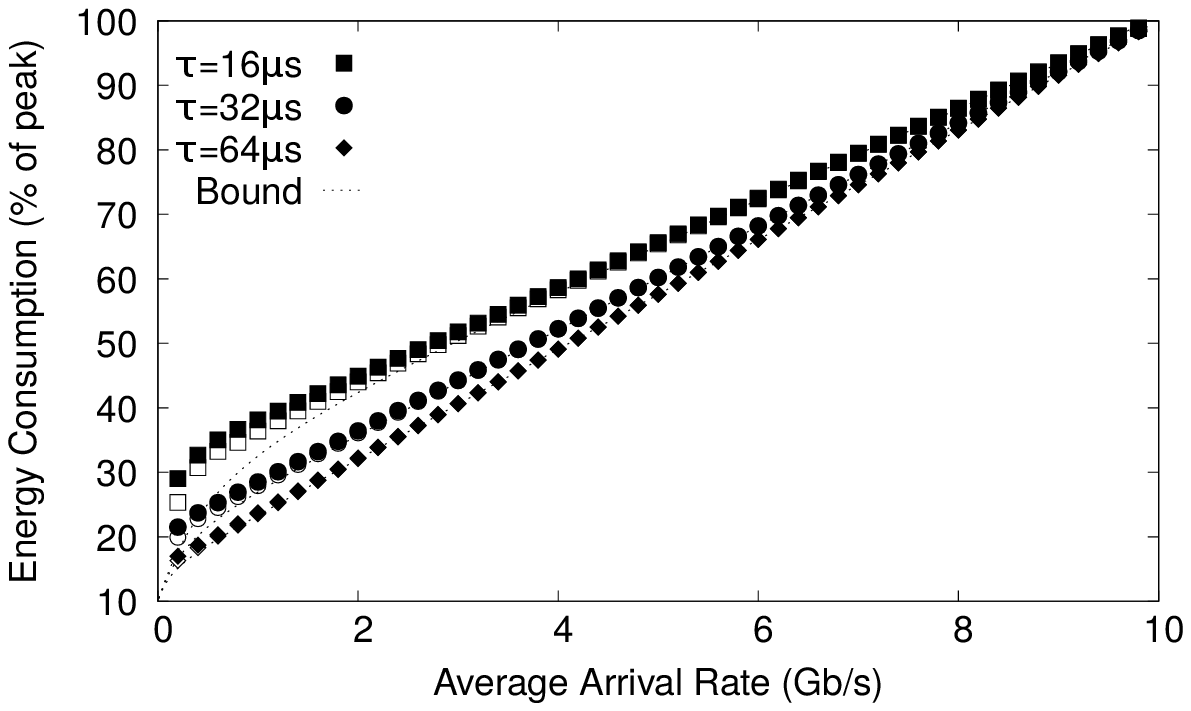}
    \label{fig:pareto-results-energy}
  }
  \subfigure[Average queuing delay.]{
    \includegraphics[width=0.85\columnwidth]{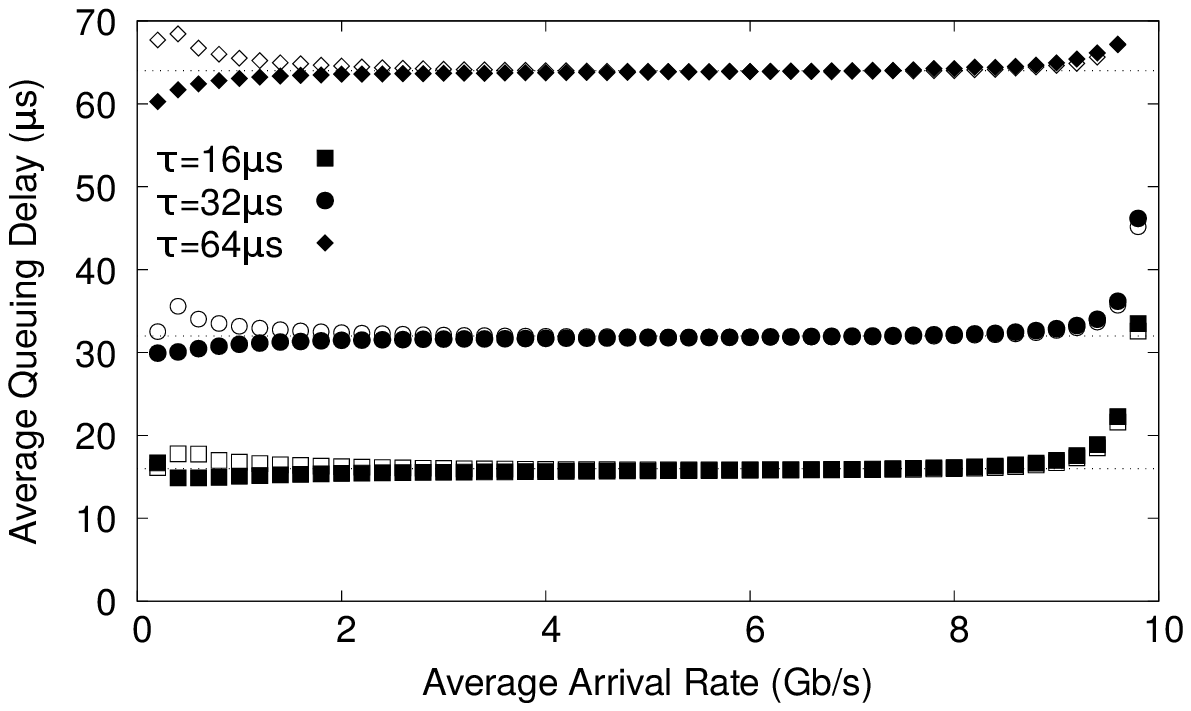}
    \label{fig:pareto-results-delay}
  }
  \caption{Simulation results with Pareto traffic. Results with
    time-based (size-based) coalescing are shown with filled
    (unfilled) points.}
  \label{fig:pareto-results}
\end{figure}

Figure~\ref{fig:pareto-results} shows that the results obtained under
these conditions are similar to those achieved with Poisson
traffic. Despite the fact that our proposals were derived from a
Poisson model, they are still able to keep the average delay near the
target value and get energy consumption very close to the lower bound
for most loads and target delays. Only at the lowest rates, the
measured delay slightly deviates from the target value due to the
effects of Pareto long-range dependence.

\begin{figure}[t]
  \centering
  \subfigure[Average timer duration (time-based coalescing).]{
    \includegraphics[width=0.85\columnwidth]{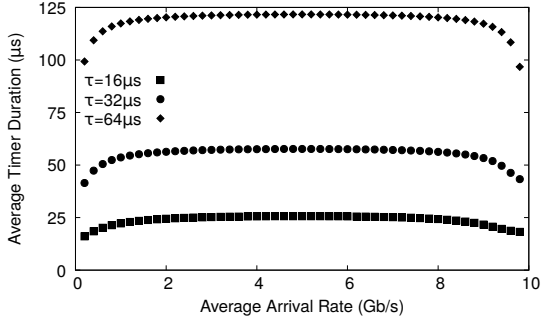}
    \label{fig:pareto-vth}
  }
  \subfigure[Average queue threshold (size-based coalescing).]{
    \includegraphics[width=0.85\columnwidth]{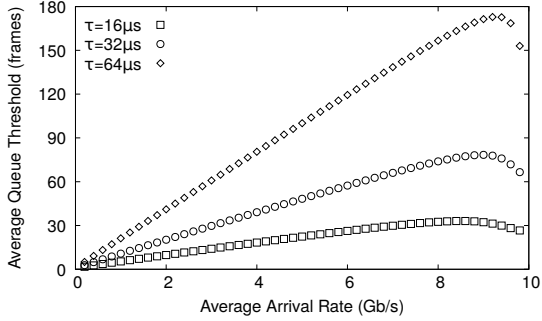}
    \label{fig:pareto-qth}
  }
  \caption{Coalescing parameters with Pareto traffic.}
  \label{fig:pareto-parameters}
\end{figure}

The average values taken by the coalescing parameters are also shown
in Fig.~\ref{fig:pareto-parameters}. As expected, the proposed
techniques appropriately tune their corresponding coalescing
parameters according to the target delay and actual traffic
conditions. Note that, except for the highest and the lowest arrival
rates, the duration of the sleeping timer with time-based coalescing
is roughly constant for a given target delay. On the other hand,
size-based coalescing selects greater queue thresholds as traffic load
increases except at the highest rates, in which the queue threshold
must be decreased to cope with the unavoidable increase on frame delay
due to excessive load. These results are in line with those obtained
analytically in~Sect.~\ref{sec:d_coalescing} (and shown in
Fig.~\ref{fig:coal-parameters}).

\subsection{Dynamic Coalescing with Real Traffic}

\begin{figure}[t]
  \centering
  \subfigure[Energy consumption.]{
    \includegraphics[width=0.85\columnwidth]{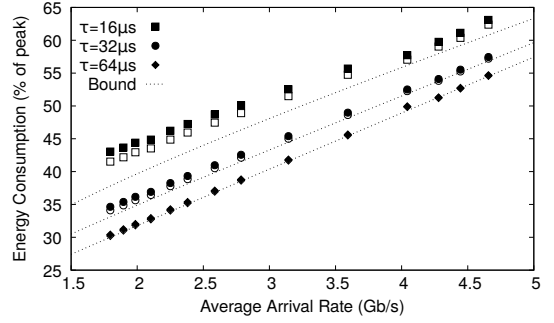}
    \label{fig:caida-results-energy}
  }
  \subfigure[Average queuing delay.]{
    \includegraphics[width=0.85\columnwidth]{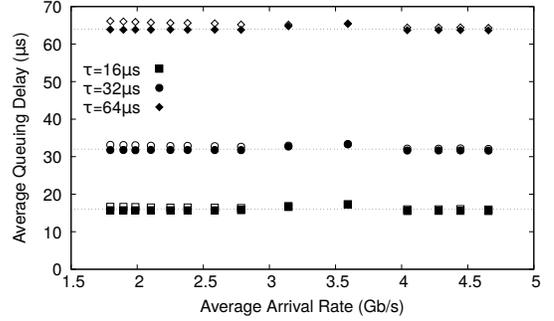}
    \label{fig:caida-results-delay}
  }
  \caption{Simulation results for CAIDA traces. Results with
    time-based (size-based) coalescing are shown with filled
    (unfilled) points.}
  \label{fig:caida-results}
\end{figure}

\begin{figure}[t]
  \centering
  \subfigure[Average timer duration (time-based coalescing).]{
    \includegraphics[width=0.85\columnwidth]{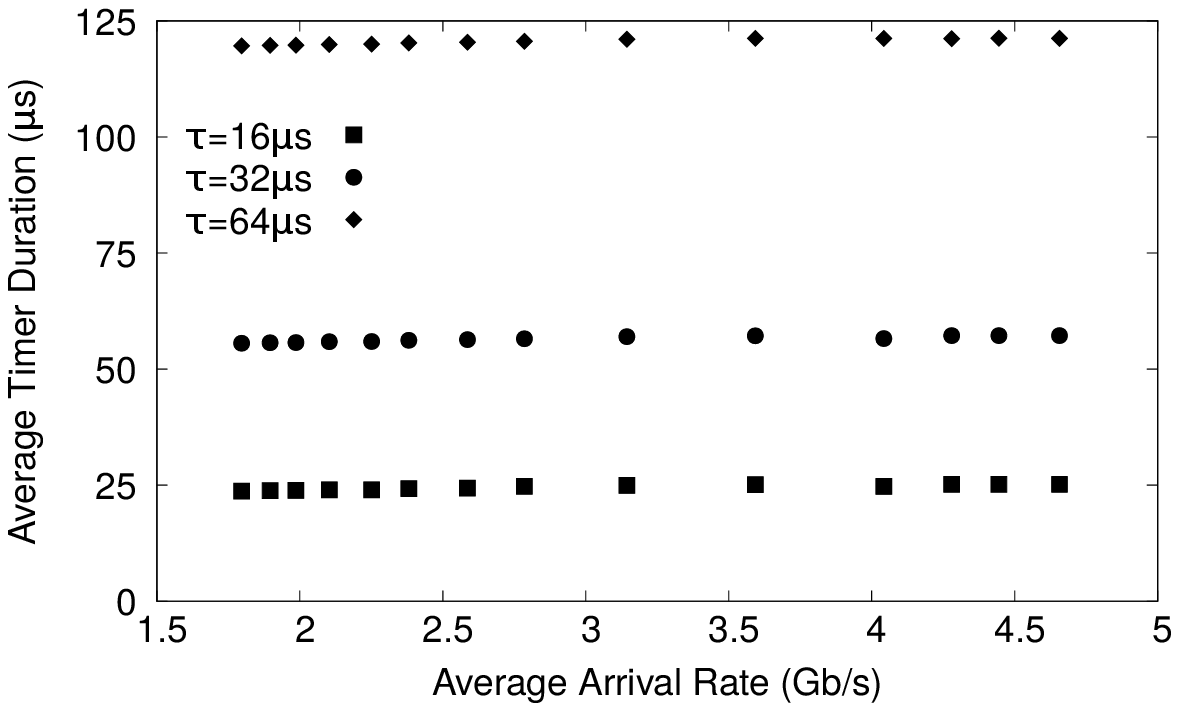}
    \label{fig:caida-vth}
  }
  \subfigure[Average queue threshold (size-based coalescing).]{
    \includegraphics[width=0.85\columnwidth]{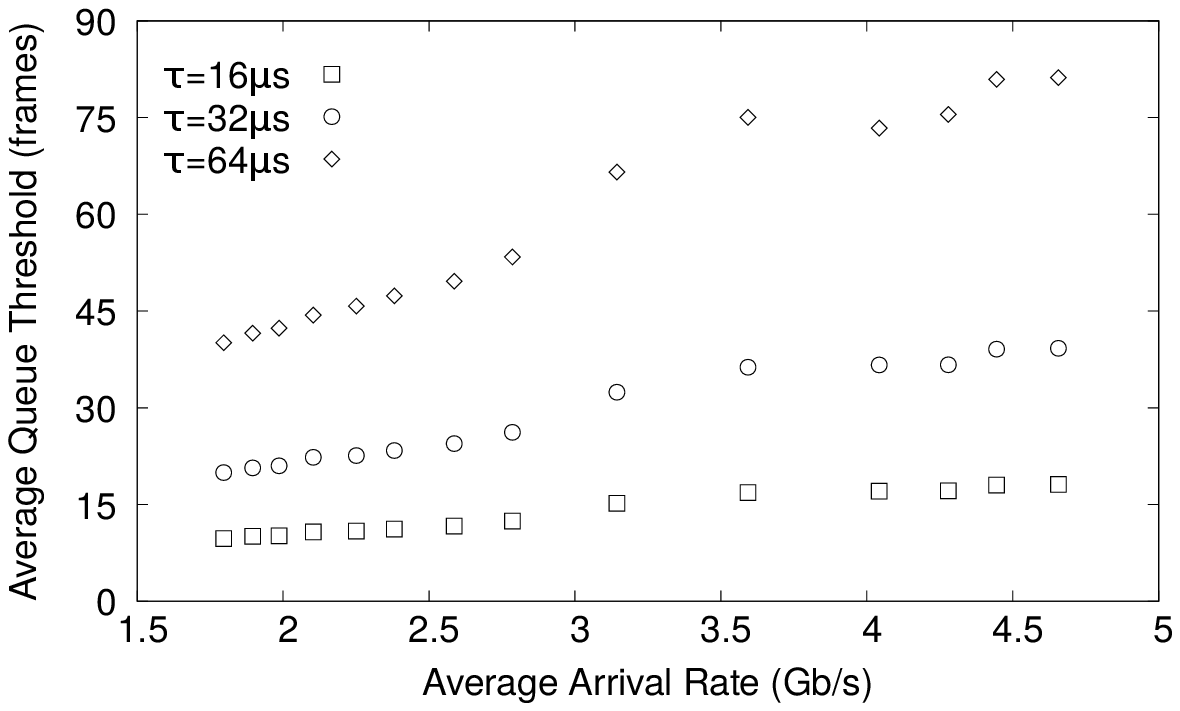}
    \label{fig:caida-qth}
  }
  \caption{Coalescing parameters for CAIDA traces.}
  \label{fig:caida-parameters}
\end{figure}

Additionally, we evaluated the proposed techniques with real world
traffic traces. In particular, we analyzed several CAIDA traces
captured on a \SI{10}{\giga b\per\s} backbone Ethernet
link~\cite{caida15}. Figure~\ref{fig:caida-results} shows the results
obtained with these traces. As in the previous experiments with
synthetic traffic, our proposals, when configured with moderate target
delays, are able to almost minimize energy consumption while keeping
the average queuing delay close to the target value. Remarkably, the
proposed techniques still work well even under these more realistic
conditions with interarrival times and frame sizes measured in a real
link since, as shown in Fig.~\ref{fig:caida-parameters}, they are able
to properly adjust the coalescing parameters according to the target
delay and actual traffic conditions.

\begin{figure}[t]
  \centering
  \subfigure[$\lambda=1.8\,$Gb/s.]{
    \includegraphics[width=0.85\columnwidth]{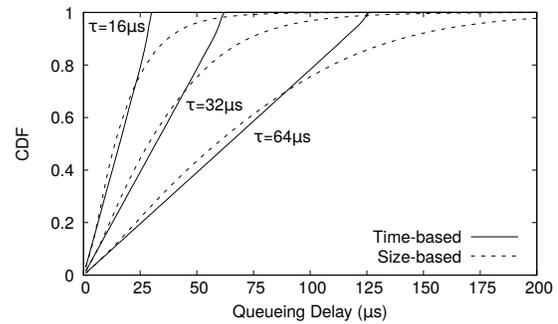}
    \label{fig:caida-cdf-134700}
  }
  \subfigure[$\lambda=3.1\,$Gb/s.]{
    \includegraphics[width=0.85\columnwidth]{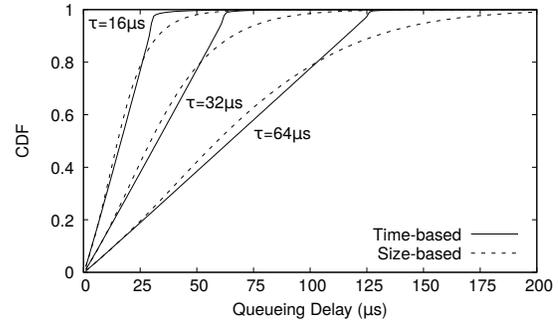}
    \label{fig:caida-cdf-134747}
  }
  \subfigure[$\lambda=4.7\,$Gb/s.]{
    \includegraphics[width=0.85\columnwidth]{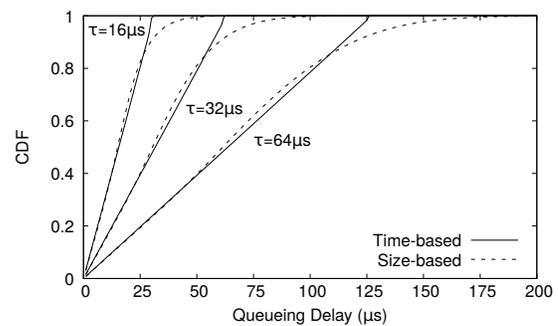}
    \label{fig:caida-cdf-130400}
  } 
  \caption{CDF of the queuing delay measured for CAIDA traces.}
  \label{fig:caida-cdf}
\end{figure}

Finally, to go deeper into the effects of coalescing on data traffic,
we computed the cumulative distribution function (CDF) of the queuing
delay in the simulated scenarios. CDFs for different arrival rates and
target delays are shown in Fig.~\ref{fig:caida-cdf}. Seemingly, the
queuing delay with time-based coalescing practically follows an
uniform distribution on the interval $[0,2\tau]$ while, with
size-based coalescing, the queuing delay appears exponentially
distributed and a significant amount of frames suffer delays greater
than~$2\tau$, especially at lower rates. For instance, if a size-based
coalescer configured with a \SI{64}{\us} target delay is applied to
the trace with $\lambda=1.8\,$Gb/s, almost~\SI{14}{\percent} of the
frames experience a delay higher than~\SI{128}{\us}, and even
\SI{3}{\percent} of them suffer delays greater
than~\SI{192}{\us}. Therefore, to avoid that some frames suffer
excessive delays, size-based coalescers should necessarily incorporate
an additional timer that limits the maximum time the interface spends
in the low power mode.

\section{Recommendations}
\label{sec:recommend}

In this section, we provide some guidelines for the selection of the
most convenient dynamic algorithm in scenarios with different delay
requirements. Following the results obtained in the previous
experiments, we recommend that, in those scenarios with flexible delay
requirements, EEE~interfaces implement dynamic time-based coalescers
with target delays in the range 32--\SI{128}{\us} so that significant
energy savings can be obtained without too much QoS~degradation.

Recall that, with target delays in this range, although the size-based
coalescing algorithm induces slightly longer sleeping periods than the
time-based one, this barely has noticeable effects on energy
consumption and both algorithms approximately consume the same amount
of energy. Consequently, we recommend using time-based coalescing
since it is easier to implement as it just requires firing a timer
while size-based coalescing requires a frame counting module to
trigger the wake-up. Moreover, it is not worth exploring new and more
advanced (and surely more complex) coalescing techniques since the
energy consumed when using time-based coalescing with target delays
from a few tens of microseconds is already close enough to the lower
bound.

Also, recall that time-based coalescing implicitly limits the maximum
frame delay to twice the target delay. In contrast, with size-base
coalescing, a significant amount of frames experience excessively long
delays. This undesirable side-effect of size-base coalescing can only
be avoided if an additional timer that limits the maximum coalescing
time is also included, that is, applying time-based coalescing at the
same time.

Finally, for those scenarios with stringent delay requirements, we
recommend using size-based coalescing configured with a target delay
lower than \SI{32}{\us} to achieve greater energy savings. Probably,
in such scenarios, a static timer that bounds the maximum queuing
delay should be also fired to avoid annoying delays.

\section{Conclusions}
\label{sec:conclusions}

This paper provides new and helpful insights on the practical
efficiency limits of dynamic coalescing techniques. We presented new
open-loop adaptive versions of both time-based and size-based
coalescing techniques that dynamically adjust the coalescing parameter
according to actual traffic conditions under the constraint of a given
average frame delay. We have also derived the fundamental limits on
the maximum energy savings when considering a target average frame
delay and compared them with the energy savings obtained when using
our proposals.

Analytical and simulation results show that the energy consumption of
both proposals greatly approximates to its fundamental limits when the
target delay is configured with values larger than just a few tens of
microseconds. Based on our experiments, we have also provided
guidelines for the selection of the most appropriate coalescing
technique in accordance with the allowable delay characteristics. In
particular, we recommend the application of the dynamic time-based
algorithm in most scenarios because of its simplicity and its ability
to bound the maximum frame delay.

As future work, we plan to research new and more energy efficient
dynamic coalescing techniques for those scenarios with very stringent
delay requirements.

\section*{Acknowledgments}

Support for CAIDA's Internet Traces is provided by the National
Science Foundation, the US Department of Homeland Security, and CAIDA
Members.

This work was supported by the European Regional Development Fund
(ERDF) and the Galician Regional Government under agreement for
funding the Atlantic Research Center for Information and Communication
Technologies (atlanTTic), and by the ``Ministerio de Economía,
Industria y Competitividad'' through the project TEC2017-85587-R of
the ``Programa Estatal de Investigación, Desarrollo e Innovación
Orientada a los Retos de la Sociedad,'' (partially financed with FEDER
funds)
    
\bibliographystyle{IEEEtran} 
\bibliography{IEEEabrv,coalescing}

\end{document}